\documentclass{article}

\usepackage{arxiv}

\usepackage[utf8]{inputenc} 
\usepackage[T1]{fontenc}    
\usepackage{hyperref}       
\usepackage{url}            
\usepackage{booktabs}       
\usepackage{amsfonts}       
\usepackage{nicefrac}       
\usepackage{microtype}      
\usepackage{lipsum}
\usepackage{graphicx}

\usepackage[american]{babel}
\usepackage{csquotes}
\usepackage{tipa}

\usepackage{amsmath}
\usepackage{amssymb}

\usepackage{apacite}

\usepackage{subcaption}

\title{Derivation of Fitts' law from the Task Dynamics model of speech production}

\author{
 Tanner Sorensen \\
  Signal Analysis and Interpretation Laboratory \\
  University of Southern California \\
  Los Angeles, CA, 90089 \\
  \texttt{tsorense@usc.edu} \\
   \And
 Adam Lammert \\
  Department of Biomedical Engineering \\
  Worcester Polytechnic Institute \\
  Worcester, MA 01609 \\
  \texttt{alammert@wpi.edu} \\
  \And
 Louis Goldstein \\
  Department of Linguistics\\
  University of Southern California\\
  Los Angeles, CA 90089 \\
  \texttt{louisgol@usc.edu} \\
  \And
 Shrikanth Narayanan \\
  Signal Analysis and Interpretation Laboratory \\
  University of Southern California\\
  Los Angeles, CA 90089 \\
  \texttt{shri@sipi.usc.edu} \\
}

\begin{document}
\maketitle
\begin{abstract}
Fitts' law is a linear equation relating movement time to an index of movement difficulty.
The recent finding that Fitts' law applies to voluntary movement of the vocal tract raises the question of whether the theory of speech production implies Fitts' law.
The present letter establishes a theoretical connection between Fitts' law and the Task Dynamics model of speech production.
We derive a variant of Fitts' law where the intercept and slope are functions of the parameters of the Task Dynamics model and the index of difficulty is a product logarithm, or Lambert W function, rather than a logarithm.
\end{abstract}


\section{Background}

The work of~\citeA{fitts1954} provided support for a linear relation between the average duration $T$ of a movement and an index of difficulty $I_d = \log(2A/W)$, where $A$ is the average movement amplitude and $W$ is the error tolerance. Thus, Fitts' law can be stated as the linear equation
\begin{equation}
\label{eq:fitts_law}
    T = \alpha + \beta I_d,
\end{equation}
where $\alpha$ is the intercept and $\beta$ is the slope.

The studies of~\citeA{lammert2018} and~\citeA{kuberski2019} were the first to provide evidence that speech movements of the vocal tract conform with Fitts' law.
\citeA{lammert2018} discovered that Fitts' law applied to vocal tract constriction movements in the coda position of the syllable (e.g., post-vocalic consonant \textipa{[k]} in the word ``pack'' \textipa{[p\super h\ae k]}), with a subset of subjects exhibiting Fitts' law for movements in other syllabic positions as well.
\citeA{kuberski2019} found that Fitts' law applied to tongue tip and tongue dorsum constriction movements in the onset position of the syllable when speech rate exceeded a subject-specific threshold.
The findings of \citeA{lammert2018} and \citeA{kuberski2019} provided evidence that Fitts' law applies to vocal tract movements just as it does to movements of the upper limb \cite{fitts1954, fitts1964}, lower limb \cite{drury1975}, and postural control \cite{duarte2005}. 
Together, these studies support Fitts' law in a variety of tasks in several domains of voluntary movement.

In addition to inspiring the speech experiments described above, Fitts' law informed the theory of speech production,
with \citeA<>[\S7, ``Speaking Rate Effects'']{guenther1995} showing that the Directions into Velocities of Articulators model of neural control of speech movements produced movements consistent with Fitts' law,
and \citeA<>[\S2 ``Background: Theoretical Framework'']{lammert2018} showing that the Task Dynamics model of speech production~\cite{saltzman1989} conformed with Fitts' law under certain conditions.

The purpose of the present letter is to establish a theoretical connection between Fitts' law and the Task Dynamics model of speech production~\cite{saltzman1989}.
Section~\ref{sec:theoretical_background} presents a derivation of the result that Fitts' law has intercept $\alpha = -1/\sqrt{k/m}$ and slope $\beta = 1/\sqrt{k/m}$, where gestural stiffness $k$ and mass $m$ are parameters of the exponential movement trajectory in the Task Dynamics model of speech production.
The index of difficulty involves the product logarithm, or Lambert W function~\cite{corless1996}, rather than the logarithm of existing indices of difficulty.
Section~\ref{sec:comparison} compares this derivation with that of \citeA{lammert2018} and shows that the present derivation is more general than that of \citeA{lammert2018} and conforms with the speech articulator movements typically analyzed in studies of vocal tract kinematics.

\section{A derivation of Fitts' law}
\label{sec:theoretical_background}

The present section derives that Fitts' law has intercept $\alpha = -1/\sqrt{k/m}$ and slope $\beta = 1/\sqrt{k/m}$, where gestural stiffness $k$ and mass $m$ are parameters of the exponential movement trajectory in the Task Dynamics model of speech production~\cite{saltzman1989}.
The derivation results in an alternative index of difficulty that involves the product logarithm, or Lambert W function~\cite{corless1996}, rather than the logarithm of Fitts' indices of difficulty~\cite{fitts1954}.
We demonstrate that the alternative index of difficulty is an index of movement difficulty that is qualitatively similar to that of \citeA{fitts1954}, as well as those of \citeA{lammert2018} and \citeA{kuberski2019} used in speech studies of Fitts' law.

The equation of motion for a gesture with stiffness $k$, damping $b$, and mass $m$ is determined by the second-order linear system
\begin{equation}
    m\ddot{z} + b\dot{z} + kz = 0, \label{eq:lsom}
\end{equation}
where the parameters satisfy the condition $b^2 - 4km = 0$ of critical damping. The fixed point is located at the origin $z = 0$.
Under these conditions, the solution of the system is the following~\cite[Chapter 1, Section 4]{jordan2007}.
\begin{equation}
    z(t) = (c_0 + c_1t) \exp{\left(-t\sqrt{k/m}\right)} \label{eq:general_solution}
\end{equation}
We solve for $c_0$ and $c_1$ in terms of initial conditions $z(0)$, $\dot{z}(0)$ to obtain $c_0 = z(0)$ and $c_1 = \dot{z}(0) + z(0)\sqrt{k/m}$.

If a movement begins at time zero, then the residual displacement $z(\tau)$ from the fixed point after time $\tau$ is $z(\tau) = (c_0 + c_1\tau) \exp{\left(-\tau \sqrt{k/m}\right)}$.
Taking the logarithm of both sides and applying the identity $\log(c_0 + c_1\tau) = \log(1 + c_1\tau / c_0) + \log(c_0)$, we derive the following.
\begin{align}
    \log(z(\tau)) &= \log\left(1 + \frac{c_1}{c_0} \tau \right) + \log(c_0) -\tau\sqrt{k/m} \\
    \Longleftrightarrow
    \tau &= \left. \left( \log\left(1 + \frac{c_1}{c_0} \tau \right) + \log\left(\frac{c_0}{z(\tau)}\right) \right) \middle/ \sqrt{k/m} \right. . \label{eq:nonzero_velocity}
\end{align}
In order to solve Equation~\ref{eq:nonzero_velocity} for $\tau$, we make the simplifying assumption that the initial velocity $\dot{z}(0)$ equals zero. Although this assumption limits the generality of the result we are deriving, the assumption is approximately correct in speech kinematic analysis, where speech movements are taken to begin when velocity surpasses a small threshold.

Setting initial velocity equal to zero implies the equation
\begin{equation}
\label{eq:zero_velocity}
    \left(- \tau \sqrt{k/m} - 1\right) \exp\left(-\tau\sqrt{k/m}\right) = -\frac{z(\tau)}{c_0},
\end{equation}
which has a solution for $\tau$. 
Substituting $x = -\tau \sqrt{k/m}$, we derive this solution as follows.
\begin{align}
    (x - 1) \exp(x - 1) &= -\frac{z(\tau)}{\mathrm{e}c_0} \\
    x &= \mathfrak{w}\left(-\frac{z(\tau)}{\mathrm{e}c_0}\right) + 1 \\
    \iff
    \tau &= -\frac{1}{\sqrt{k/m}} \mathfrak{w}\left(-\frac{z(\tau)}{\mathrm{e}c_0}\right) -\frac{1}{\sqrt{k/m}}
    \label{eq:solution}
\end{align}
In the above solution, $\mathrm{e}$ is Euler's number and $\mathfrak{w}$ denotes the product logarithm, also known as the Lambert W function~\cite{corless1996}.
The product logarithm is defined as the inverse of the function $f(z) = z \exp (z)$.
Although the relation defined by this definition is multivalued and has the set of complex numbers as its domain, below we introduce restrictions on the domain and codomain such that the product logarithm is a function whose domain is the real numbers.
Note that the solution to Equation~\ref{eq:solution} above follows from the identity $\mathfrak{w}((x - 1)\exp(x - 1)) = x - 1$ of the product logarithm.

\begin{figure}
\begin{subfigure}[t]{0.35\linewidth}
    \centering\includegraphics[width=\textwidth]{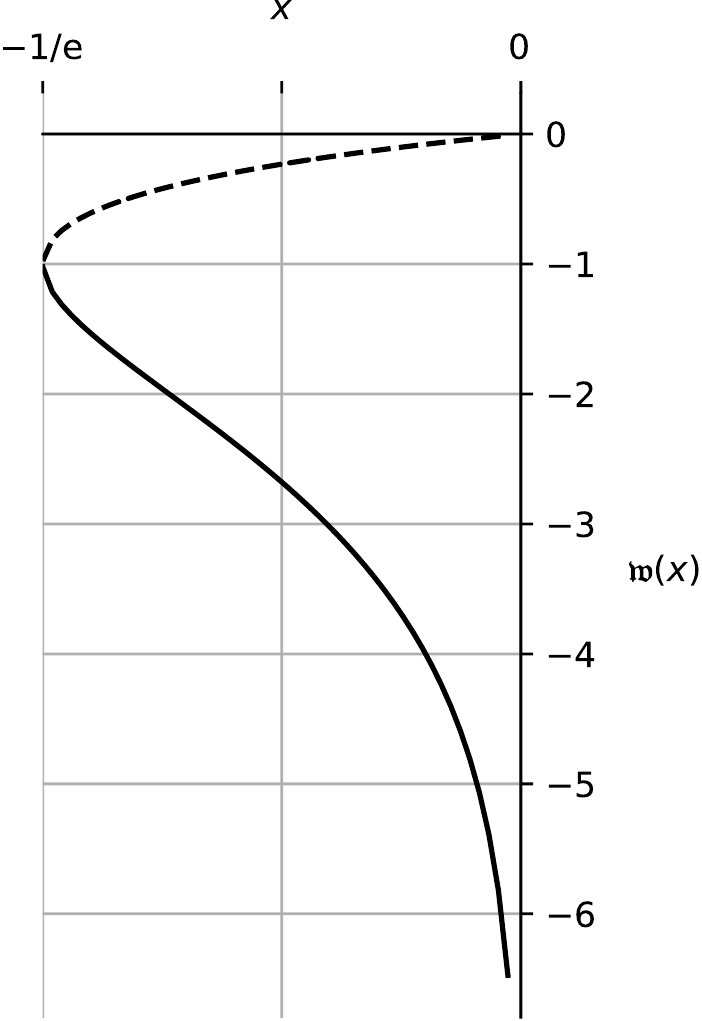}
    \caption{}
    \label{fig:lambert_w}
  \end{subfigure}\quad
  \begin{subfigure}[t]{0.6\linewidth}
    \centering\includegraphics[width=\textwidth]{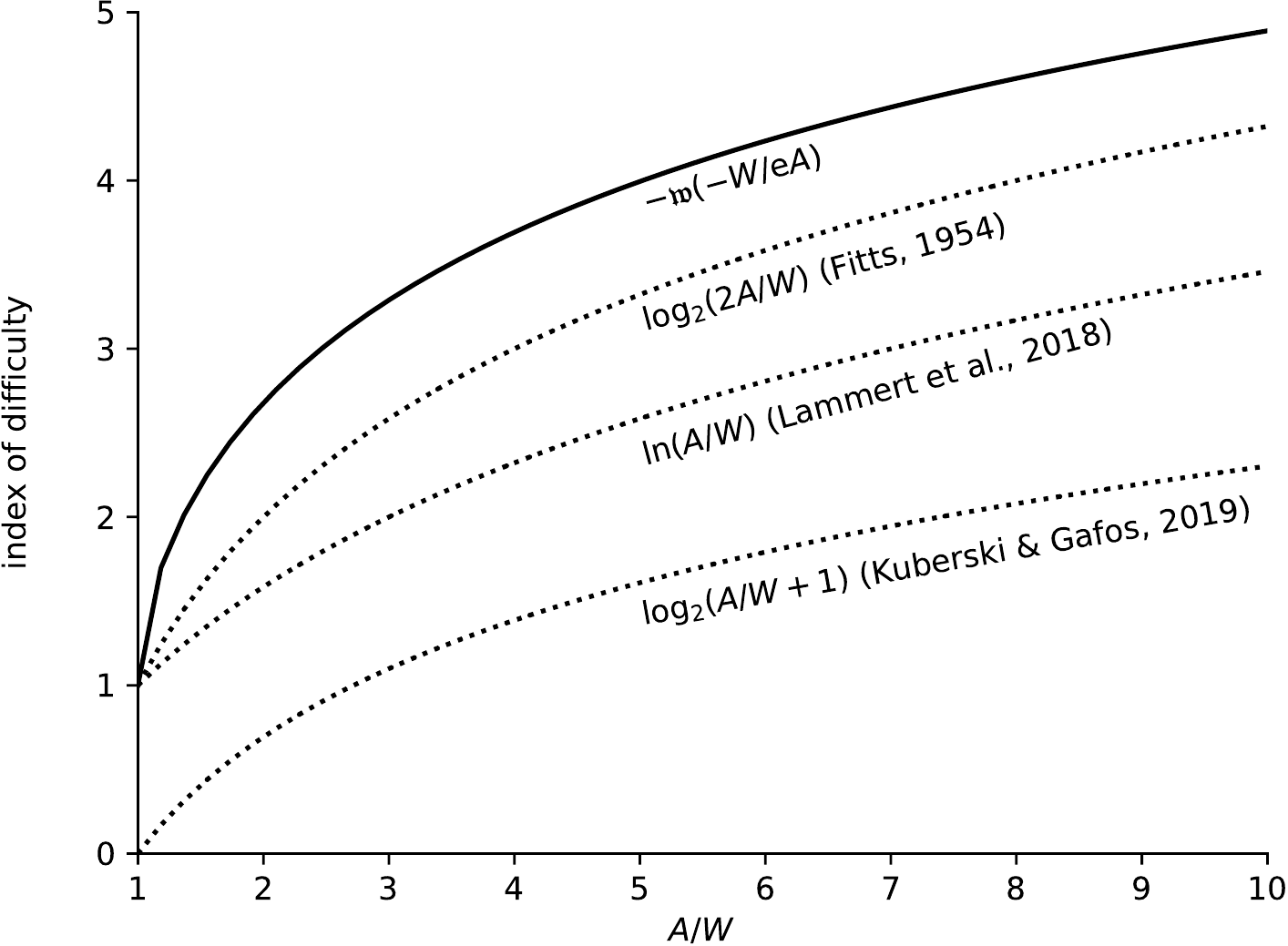}%
    \caption{}
    \label{fig:id_comparison}
  \end{subfigure}%
    \caption{\subref{fig:lambert_w}
    Graph of the real branches of the product logarithm for $x \in (-1/\mathrm{e}, 0)$. The principal real branch $\mathfrak{w}_0$ is defined as the upper branch with $\mathfrak{w} \geq -1$ (dashed)  and the real branch $\mathfrak{w}_{-1}$ is defined as the lower branch with $\mathfrak{w} < -1$ (solid).
    \subref{fig:id_comparison}
    Comparison of the index of difficulty $-\mathfrak{w}_{-1}(-W/\mathrm{e}A)$ proposed in the present study with others from the literature. The comparison is performed for different values of the ratio $A/W$, where the value $A/W = 1$ is a lower bound on possible movements.
    }
\end{figure}

Equating time $\tau$ with movement time $T$, constant $c_0$ with movement amplitude $A$, and residual displacement $z(\tau)$ with error tolerance $W$,
we derive that
\begin{equation}
    T = \alpha + \beta I'_d, \label{eq:my_fitts_law}
\end{equation}
where intercept $\alpha = -1/\sqrt{k/m}$, slope $\beta = 1/\sqrt{k/m}$, and index of difficulty $I'_d = -\mathfrak{w}(-W/\mathrm{e}A)$. Prime indicates that this index of difficulty is not identical to that of Fitts.
As movement amplitude $A$ must be greater than error tolerance $W$, we have that the domain of the product logarithm is a real number in the open interval $(-1/\mathrm{e}, 0) \subset \mathbb{R}$.
In this interval, the product logarithm has two real branches (cf. Figure~\ref{fig:lambert_w}): an upper branch $\mathfrak{w}_{0}$ (dashed) and a lower branch $\mathfrak{w}_{-1}$ (solid).
Branch $-\mathfrak{w}_{0}(-W/\mathrm{e}A)$ is monotonically decreasing in $A/W$, while branch $-\mathfrak{w}_{-1}(-W/\mathrm{e}A)$ is monotonically increasing in $A/W$.
Restricting attention to the lower branch $\mathfrak{w}_{-1}$, we find that the index of difficulty increases with movement amplitude and decreases with error tolerance.
The graph of index of difficulty as a function of $A/W > 1$ is similar in form to that of \citeA{fitts1954}, taking the value $I'_d = 1$ at $A/W = 1$ and increasing monotonically with $A/W$ (cf. Figure~\ref{fig:id_comparison}, solid).
Furthermore, movement time is a linear function of the index of difficulty (cf. Equation~\ref{eq:my_fitts_law}).
These properties make the prediction of Equation~\ref{eq:my_fitts_law} qualitatively similar to the prediction of Fitts' law,
the difference being the index of difficulty assumed: $I'_d = -\mathfrak{w}_{-1}(-W/\mathrm{e}A)$ under Equation~\ref{eq:my_fitts_law}; or $I_d = -\log_2(W/2A)$ under Fitts' law.

\section{Comparison with the existing kinematic derivations}
\label{sec:comparison}

Equation~\ref{eq:my_fitts_law} is not the only version of Fitts' law that has been derived from the Task Dynamics model of speech production.
\citeA{lammert2018} derived a version of Fitts' law by considering the particular solution
\begin{equation}
    z(t) = c_0 \exp\left( - t \sqrt{k/m} \right) \label{eq:particular_solution}
\end{equation}
to the critically damped linear second-order system of Equation~\ref{eq:lsom}, where initial position $c_0 = z(0)$ and initial velocity $\dot{z}(t) = -z(0)\sqrt{k/m}$.
By considering this particular solution in place of the general solution of Equation~\ref{eq:general_solution}, the solution to Equation~\ref{eq:nonzero_velocity} simplifies from a product logarithm to a logarithm, since the constant $c_1 = \dot{z}(t) + z(0)\sqrt{k/m}$ equals zero.
\begin{align}
    \tau &= \left. \left( \log\left(1 + \frac{c_1}{c_0} \tau \right) + \log\left(\frac{c_0}{z(\tau)}\right) \right) \middle/ \sqrt{k/m} \right. . \tag{\ref{eq:nonzero_velocity} revisited} \\
    &= \left. \log\left(\frac{c_0}{z(\tau)} \right) \middle/ \sqrt{k/m} \right. .
\end{align}
Substituting movement time $T$ for time $\tau$, movement amplitude $A$ for constant $c_0$, and error tolerance $W$ for residual displacement $z(\tau)$, we derive the version of Fitts' law obtained by \citeA<>[Section ``Background: Theoretical framework'', Equation 4]{lammert2018}.
\begin{equation}
    T = \frac{1}{\sqrt{k/m}} \log\left( \frac{A}{W} \right) \label{eq:lammert_ID}
\end{equation}
\citeA{lammert2018} showed the conditions under which this equation is equivalent to the original Fitts' law~\cite{fitts1954}.

\begin{figure}
    \centering
    \includegraphics[width=\textwidth]{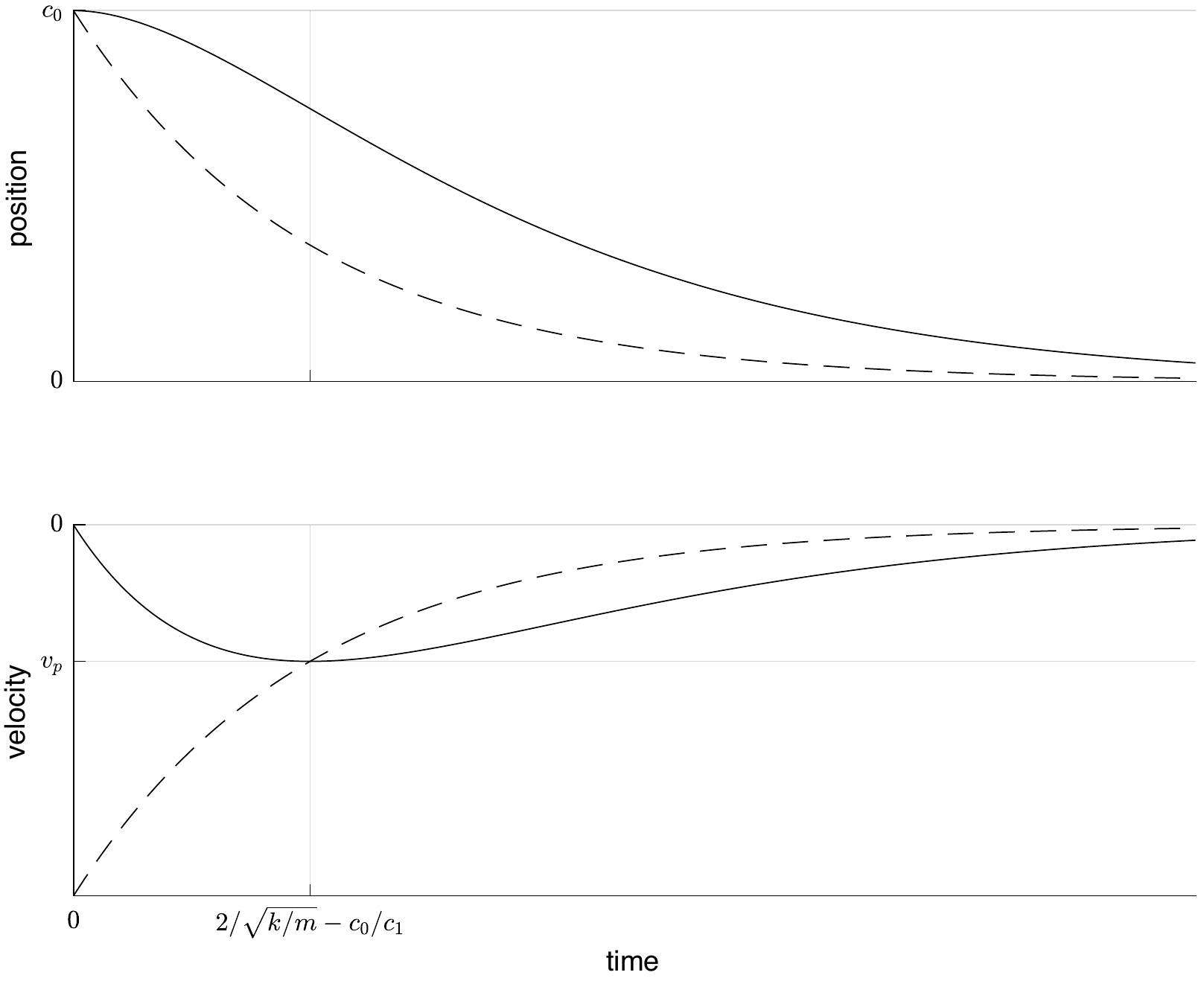}
    \caption{Comparison of solution with initial velocity set to zero (solid) and with initial velocity set to $-z(0)\sqrt{k/m}$ (dashed). With zero initial velocity (solid), peak velocity (horizontal line, bottom panel) occurs at time $2/\sqrt{k/m} - c_0/c_1$ (vertical line), where $k$ is gestural stiffness, $m$ is virtual mass, $c_0$ is the initial position $z(0)$, and $c_1 = \dot{z}(0) - z(0) \sqrt{k/m}$. With initial velocity $-z(0)\sqrt{k/m}$ (dashed), peak velocity occurs at time zero (i.e., is simultaneous with the very start of the movement).}
    \label{fig:kinematics}
\end{figure}

The reason to prefer the present index of difficulty over the logarithmic index of difficulty of Equation~\ref{eq:lammert_ID} is that the derivation of the present index of difficulty considers the general solution to the critically damped second-order linear system of Equation~\ref{eq:lsom} with initial velocity equal to zero.
In contrast, the derivation of Equation~\ref{eq:lammert_ID} replaces the general solution of Equation~\ref{eq:general_solution} with the particular solution $z(t) = c_0 \exp(-t \sqrt{k/m})$ or, equivalently, considers the general solution of Equation~\ref{eq:general_solution} with the particular nonzero initial velocity $-z(0)\sqrt{k/m}$.
The condition of zero initial velocity in the present derivation is approximately correct in speech kinematic analysis, where speech movements are taken to begin when velocity surpasses a small threshold.
Moreover, replacing the general solution of Equation~\ref{eq:general_solution} with the particular solution $z(t) = c_0 \exp(-t \sqrt{k/m})$ implies that the initial velocity $-z(0)\sqrt{k/m}$ is the peak velocity of the movement (cf. Figure~\ref{fig:kinematics}, dashed movement traces).
The implication that the initial velocity equals the peak velocity does not agree with the observation that peak velocity occurs not at the very start of a movement, but rather closer to the temporal midpoint of a movement~\cite{munhall1985, ostry1987, byrd1998, sorensen2015}.

\section{Acknowledgments}
The authors acknowledge funding through National
Institutes of Health (NIH) Grant Nos. R01DC007124 and
T32DC009975, and National Science Foundation (NSF)
Grant No. 1514544. The content of this paper is solely the
responsibility of the authors and does not necessarily
represent the official views of the NIH or NSF.

\bibliographystyle{apacite}  


\end{document}